\documentclass[fleqn,twoside]{article}
\usepackage[headings]{espcrc2}
\readRCS
$Id: espcrc2.tex,v 1.2 2004/02/24 11:22:11 spepping Exp $
\usepackage{graphicx}
\usepackage[figuresright]{rotating}

\newcommand{\AmS}{{\protect\the\textfont2
  A\kern-.1667em\lower.5ex\hbox{M}\kern-.125emS}}
\title{Effects of density and parametrization on scattering observables
}

\author{M. Bhuyan\address[SU]{School of Physics, Sambalpur University, Jyotivihar-768 019, India}%
and
S. K. Patra\address[IOP]{Institute of Physics, Sachivalaya Marg, Bhubaneswar-751 005, India }}%

\runauthor{M. Bhuyan and S.K. Patra}

\begin{document}

\begin{abstract}

We calculate the density distribution of protons and neutrons for
$^{40,42, 44,48}Ca$ in the frame-work of relativistic 
mean field (RMF) theory with NL3 and G2 parameter sets. The microscopic 
proton-nucleus optical potential for $p+^{40}Ca$ system is evaluted from
Dirac NN-scattering amplitude and the density of the target nucleus using
Relativistic-Love-Franey and McNeil-Ray-Wallace parametrizations.
Then we estimate the scattering observables, such as elastic differential 
scattering cross-section, analysing power and the spin observables with 
relativistic impulse approximation. We compare the results with the 
experimental data for some selective cases and found that the use of density 
as well as the scattering matrix parametrization is crucial for the 
theoretical prediction.
\vspace{1pc}
\end{abstract}
\maketitle

Explaining the nuclear structure by taking the tool of nuclear reaction 
is one of the most curious and challenging solution for Nuclear Physics 
both in theory and laboratory. So far the elastic scattering reaction of 
Neucleon-Nucleus is more interesting than that of Nucleus-Nucleus at 
laboratory energy $E_{lab} \simeq$ 1000 MeV. The Neucleon-Nucleus interaction 
provides a fruitful source to determine the nuclear structure and a clear path 
toward the formation of exotic nuclei in laboratory. One of the theoritical 
method to study such type of reaction is the Relativistic Impulse Approximation 
(RIA). In a wide range of energy interval, the conventional impulse approximation
\cite{fedd63,mahu78} reproduces quantitatively the main features of 
quasi-elastic scattering for medium mass nuclei \cite{bala68,glau55}. The 
observables of the elastic scattering
reaction  not only depend on the energy of the incident particle but also 
on the kinematic parameter as well as the density discributions of the
target nucleus.  In the present letter, our motivation is to calculate the
nucleon-nucleus elastic differential scattering cross-section 
($\frac{d\sigma}{d\Omega}$) and other quantities, like optical potential 
($U_{opt}$), analysing power ($A_y$) and spin observables ($Q-$value) 
taking input as relativistic mean field (RMF) and recently proposed 
effective field theory motivated relativistic mean field (E-RMF)
density. The RMF and E-RMF densities are obtained from the most successful 
NL3 \cite{lala97} and advanced G2 \cite{tang96} parameter sets, respectively. 
As representative cases, we used these target densities folded with the 
NN-aplitude of 1000 MeV energetic proton projectile with Relativistic-Love-Franey 
(RLF) and McNeil-Ray-Wallace (MRW) parametrizations \cite{neil83} for 
$^{40,42,44,48}Ca$ in our calculations. 

The RMF and E-RMF theories are well documented \cite{tang96,patra01a,patra91}
 and for completeness we outline here very briefly the formalisms for finite nuclei.
The energy density functional of the E-RMF model for finite nuclei is written 
as \cite{ser97,fur96},

\begin{eqnarray}
\mathcal{E}(\mathbf{r})  =  \sum_\alpha \varphi_\alpha^\dagger \Bigg\{ -i %
\mbox{\boldmath$\alpha$} \!\cdot\! \mbox{\boldmath$\nabla$} + \beta (M -
\Phi) + W + \nonumber \\ 
\frac{1}{2}\tau_3 R + \frac{1+\tau_3}{2} A 
- \frac{i}{2M} \beta \mbox{\boldmath$\alpha$}\!\cdot\! (f_v %
\mbox{\boldmath$\nabla$} W + \frac{1}{2}f_\rho\tau_3 \mbox{\boldmath$\nabla$}
\nonumber \\
R + \lambda \mbox{\boldmath$\nabla$} A ) + \frac{1}{2M^2}\left
(\beta_s + \beta_v \tau_3 \right ) \Delta A \Bigg\} \varphi_\alpha  \nonumber
\\
 \null + \left ( \frac{1}{2} + \frac{\kappa_3}{3!}\frac{\Phi}{M} + \frac{%
\kappa_4}{4!}\frac{\Phi^2}{M^2}\right ) \frac{m_{s}^2}{g_{s}^2} \Phi^2 -
\frac{\zeta_0}{4!} \frac{1}{ g_{v}^2 } W^4  \nonumber \\[3mm]
\null + \frac{1}{2g_{s}^2}\left( 1 + \alpha_1\frac{\Phi}{M}\right)
\left( \mbox{\boldmath $\nabla$}\Phi\right)^2 - \frac{1}{2g_{v}^2}\left( 1
+\alpha_2\frac{\Phi}{M}\right) 
\nonumber \\
\left( \mbox{\boldmath $\nabla$}
W \right)^2
\null - \frac{1}{2}\left(1 + \eta_1 \frac{\Phi}{M} + \frac{\eta_2}{2}
\frac{\Phi^2 }{M^2} \right) \frac{{m_{v}}^2}{{g_{v}}^2} W^2 - 
\frac{1}{%
2g_\rho^2} 
\nonumber \\
\left( \mbox{\boldmath $\nabla$} R\right)^2 - \frac{1}{2} \left(
1 + \eta_\rho \frac{\Phi}{M} \right) \frac{m_\rho^2}{g_\rho^2} R^2  \nonumber
\\
\null - \frac{1}{2e^2}\left( \mbox{\boldmath $\nabla$} A\right)^2 +
\frac{1}{3g_\gamma g_{v}}A \Delta W + \frac{1}{g_\gamma g_\rho}A \Delta R ,
\end{eqnarray}
where the index $\alpha$ runs over all occupied states $\varphi_\alpha (%
\mathbf{r})$ of the positive energy spectrum, $\Phi \equiv g_{s} \phi_0(%
\mathbf{r})$, $W \equiv g_{v} V_0(\mathbf{r})$, $R \equiv g_{\rho}b_0(%
\mathbf{r})$ and $A \equiv e A_0(\mathbf{r})$.
The terms with $g_\gamma$, $\lambda$, $\beta_{s}$ and $\beta_{v}$ take care of the effects
related with the electromagnetic structure of the pion and the nucleon (see Ref.\ \cite{fur96}).
The energy density contains tensor couplings, and scalar-vector and vector-vector meson interactions,
in addition to the standard scalar self interactions $\kappa_{3}$ and $\kappa_{4}$. Thus, the E-RMF
formalism can be interpreted as a covariant formulation of density functional theory as it contains
all the higher order terms in the Lagrangian, obtained by expanding it in powers of the meson fields.
The terms in the Lagrangian are kept finite by adjusting the parameters. Further insight into the
concepts of the E-RMF model can be obtained from Ref. \cite{fur96}. It may be noted that the standard
RMF Lagrangian is obtained from that of the E-RMF by ignoring the vector-vector and scalar-vector cross
interactions, and hence does not need a separate discussion.
In each of the two formalisms (E-RMF and RMF), the set of coupled equations are solved numerically by
a self-consistent iteration method and the baryon, scalar, isovector, proton, neutron and tensor 
densities are calculated.

The numerical procedure of calculation and the detailed equations for the ground 
state properties of finite nuclei, we refere the reader to Refs. 
\cite{patra91,patra01a}. The densities obtained from  RMF (NL3) \cite{lala97} 
and E-RMF (G2) \cite{tang96} are used for folding with the NN-sacttering amplitude 
at $E_{lab}=1000 MeV$, which gives the proton-nucleus complex optical potential 
for RMF and E-RMF formalisms. RIA involves mainly two steps \cite{furn87,pard83}
of calculations for the evaluation of the NN-scattering amplitude. In this case, 
five Lorentz covariant function \cite{neil83} multiply with the so called Fermi 
invariant Dirac matrix (NN-scattering amplitudes). This NN-amplitudes are folded 
with the target densities of protons and neutrons to produced a first order 
complex optical potential $U_{opt}$. The invariant NN-scattering operater 
${\cal F}$ can be written in terms of five complex functions (the five terms 
involves in the proton-proton pp and neutron-neutron pn scattering) as follows:
\begin{eqnarray}
{\cal F(q,E)}={\cal F}^{S}
+{\cal F}^{V}\gamma^{\mu}_{(0)}\gamma_{(1)\mu}
+{\cal F}^{PS}\gamma^{5}_{(0)}\gamma^{5}_{(1)}\nonumber\\
+{\cal F}^{T}\sigma^{\mu\nu}_{(0)}\sigma_{(1)\mu\nu}
+{\cal F}^{A}\gamma^{5}_{(0)}\gamma^{\mu}_{(0)}\gamma^{5}_{(1)}\gamma_{(1)\mu},
\end{eqnarray}
where (0) and (1) are the incident and struck nucleons respectively. 
The amplitude for each ${\cal F}^{L}$ is a complex function of the Lorentz 
invariants {\it T} and {\it S} with ${\it E}=E_{lab}$ and {\it q}
is the  four momentum. We recommend the redears for detail expressions 
to Refs. \cite{bunu10,pery86,fox89,dock87,bri77,ser84,mann84,har86,ser87}.
Then the Dirac optical potential ${\it U}_{opt}(q, E)$ can be written as,
\begin{eqnarray}
{\it U}_{opt}(q, E) = \frac{-4\pi ip}{M}\langle\psi\vert
\sum_{n=1}^{A}exp^{iq.x(n)}{\cal F}(q, E; n)\vert\psi\rangle,
\end{eqnarray}
where ${\cal F}$ is the scattering operator, ${\it p}$ is the momentum of 
the projectile in the nucleon-nucleus center of mass frame, 
$\vert\psi\rangle$ is the nuclear ground state wave function for A-particle.
Finally using the Numerov algorithm the obtained wave function is matched
with the coulomb scattering solution for a boundary condition at 
$r\rightarrow \infty$ and we get the scattering observables from the
scattering amplitude, which are defined as:
\begin{eqnarray}
\frac{d\sigma}{d\Omega}\equiv\vert A(\theta)\vert ^{2}+\vert B(\theta)\vert ^{2}\\
A_{y}\equiv\frac{2Re[A^{*}(\theta)B(\theta)]}{d\sigma /d\Omega}\\
and\nonumber\\
Q\equiv\frac{2Im[A(\theta)B^{*}(\theta)]}{d\sigma /d\Omega}.
\end{eqnarray}

Now we present our calculated results of neutrons and protons density
distribution obtained from the RMF and E-RMF formalisms \cite{patra01a}.
Then we evaluate the scattering observables using these densities in 
the relativistic impulse approximation, which involves the following 
two steps: in the first step we generate the complex NN-interaction
from the Lorentz invariant matrix ${\cal F}^L(q,E)$ as defined in
Eq. (2). Then the interaction is folded with the ground state target 
nuclear density for both the RLF and MRW parameters \cite{neil83} 
separately and obtained the nucleon-nucleus complex optical potential 
$U_{opt}(q,E)$ for the parametrisations. It is to be noted that pairing 
interaction is taken care using the Pauli blocking approximation. 
In the second step, we solve the wave function of the scattering state
utilising the optical potential prepared in the first step by well 
known Numerov algorithm \cite{koon86}. 
The result approxumated with the non-relativistic Coulomb scattering for 
a longer range of radial component which results the scattering amplitude 
and other observables \cite{thy68}. In thr present paper we calculate the 
density distribution of protons and neutrons for $^{40,42,44,48}$Ca in 
NL3 and G2 parameter sets. From the density we evalute the optical 
potential and other scattering observables and some representative 
cases are presented in Figures $1-3$.

\begin{figure}[ht]
\begin{center}
\includegraphics[width=1.0\columnwidth]{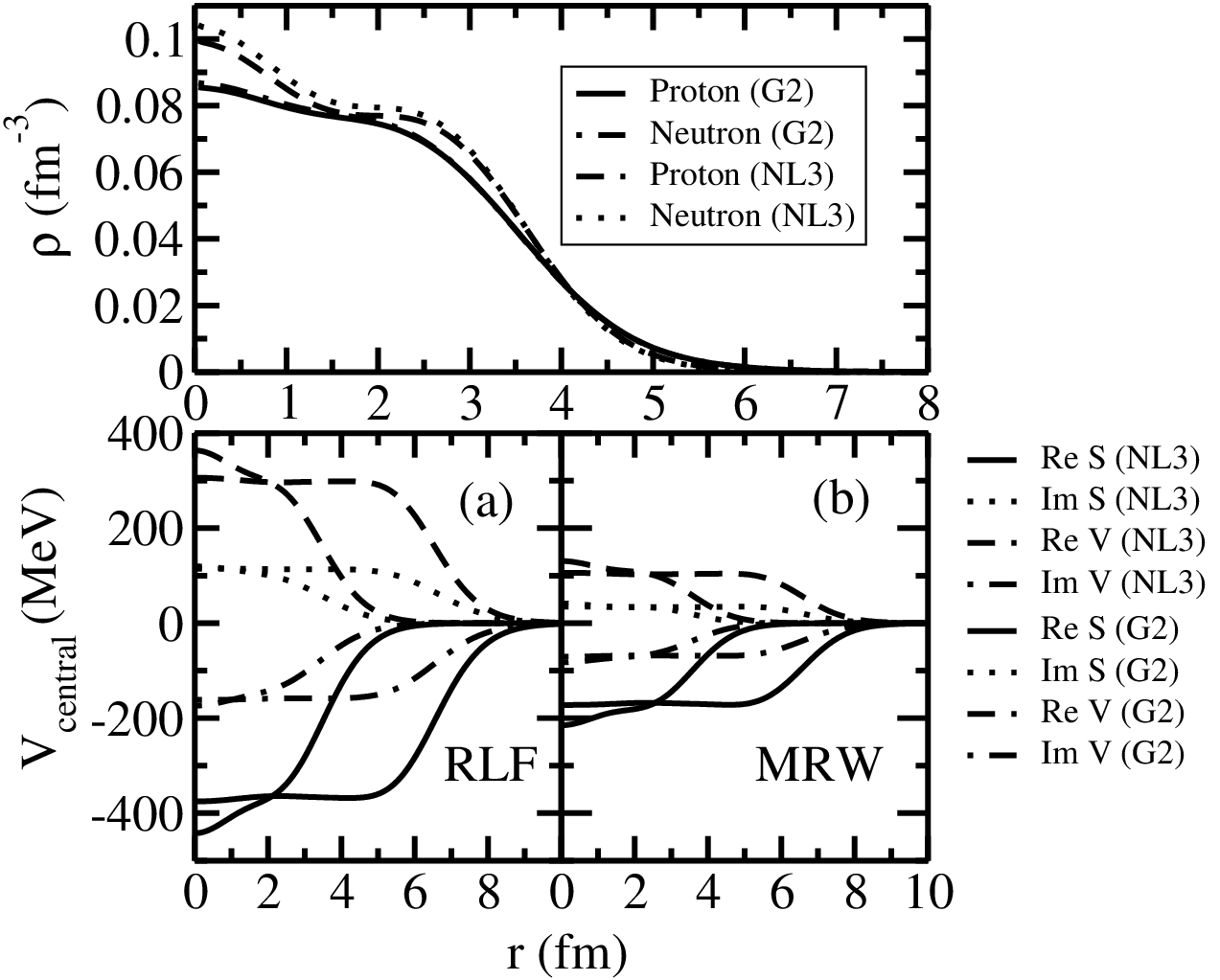}
\caption{{\it (upper panel): The neutrons and protons density
distribution for $^{40}Ca$ with NL3 and G2 parameter sets. (lower panel)
(a) the Dirac optical potential for $p+^{40}Ca$ system using RMF (NL3)
and E-RMF (G2) densities with RLF parametrisation, (b) same as (a), but
for MRW parametrisation. The projectile proton with $E_{lab}=1000$ MeV
is taken.}}
\end{center}
\end{figure}

In Fig. 1, the protons and neutrons density distribution for $^{40}Ca$ 
using NL3 and G2 parameter sets (upper panel) and the optical potential 
obtained with RLF and MRW parametrisation for $p+^{40}Ca$ at 1000 MeV 
proton energy (lower panel) are shown. From the figure, it is noticed
that, there is no significant difference in desities for RMF and E-RMF 
parameter sets. However, a careful inspection
shows a small enhancement in central density (0-1.6 fm) for NL3 set. On 
the otherhand the densities obtained from G2 elongated to a larger 
distance towards the tail part of the density distribution. 
As the optical potential is a complex function which constitute both
real and imaginary part for both scalar and vector, we have displyed
those values in the lower panel of Fig. 1. Unlike to the (upper panel) 
of protons and neutrons density distribution, here we find a large
difference of $U_{opt}(q,E)$ between the RLF and  MRW parametrisation. 
Further, the $U_{opt}(q,E)$ value of  either RLF or MRW differs 
significantly depending on the NL3 or G2 force parameters. That means, 
the optical potential not only sensitive to RLF or  MRW but also to 
the use of NL3 or G2 parameter sets.
Investigating the figure it is clear that, the extrimum magnitude
of real and imaginary part of the scalar potential are -442.2 
and 113.6 MeV for RLF (G2) and -372.4 and 109.1 MeV for RLF (NL3).
The same values for the MRW parametrisation are -219.8 and 32.8 MeV 
with G2 and -175.1 and 33.2 MeV with NL3 sets. In case of vector 
potential, the extremum values for real and imaginary parts are 361.3 
and -179.2 MeV for RLF (G2) and 279.2 and -164.8 MeV for RLF (NL3)
but with MRW parametrisation these are appeared at 128.1 and 
-87.4 MeV in G2 and 99.2 and -76.6 MeV in NL3. From these large
variation in magnitude of scalar and vector potentials,  
it is clear that the predicted results not only depend on the
input target density, but also highly sensitive with the kinematic
of the reaction dynamics. A further analysis of the results for
the optical potential with NL3 and G2, it suggest that the 
$U_{opt}$ value extends for a larger distance in NL3 than G2.
For example, with RLF the central part of $U_{opt}$ with G2 is more 
expanded than with NL3 and ended at $r\sim 6 fm$, whereas
the optical potential persists till $r\sim 8 fm$ in NL3. 
Similar situation is also valid in MRW parametrisation. This
nature of the potential suggests the applicability of NL3 over
G2 force parameter. This is because in case of NL3 the soft-core
interaction between the projectile and the target nucleon is
more effective. 

\begin{figure}[ht]
\begin{center}
\includegraphics[width=1.0\columnwidth]{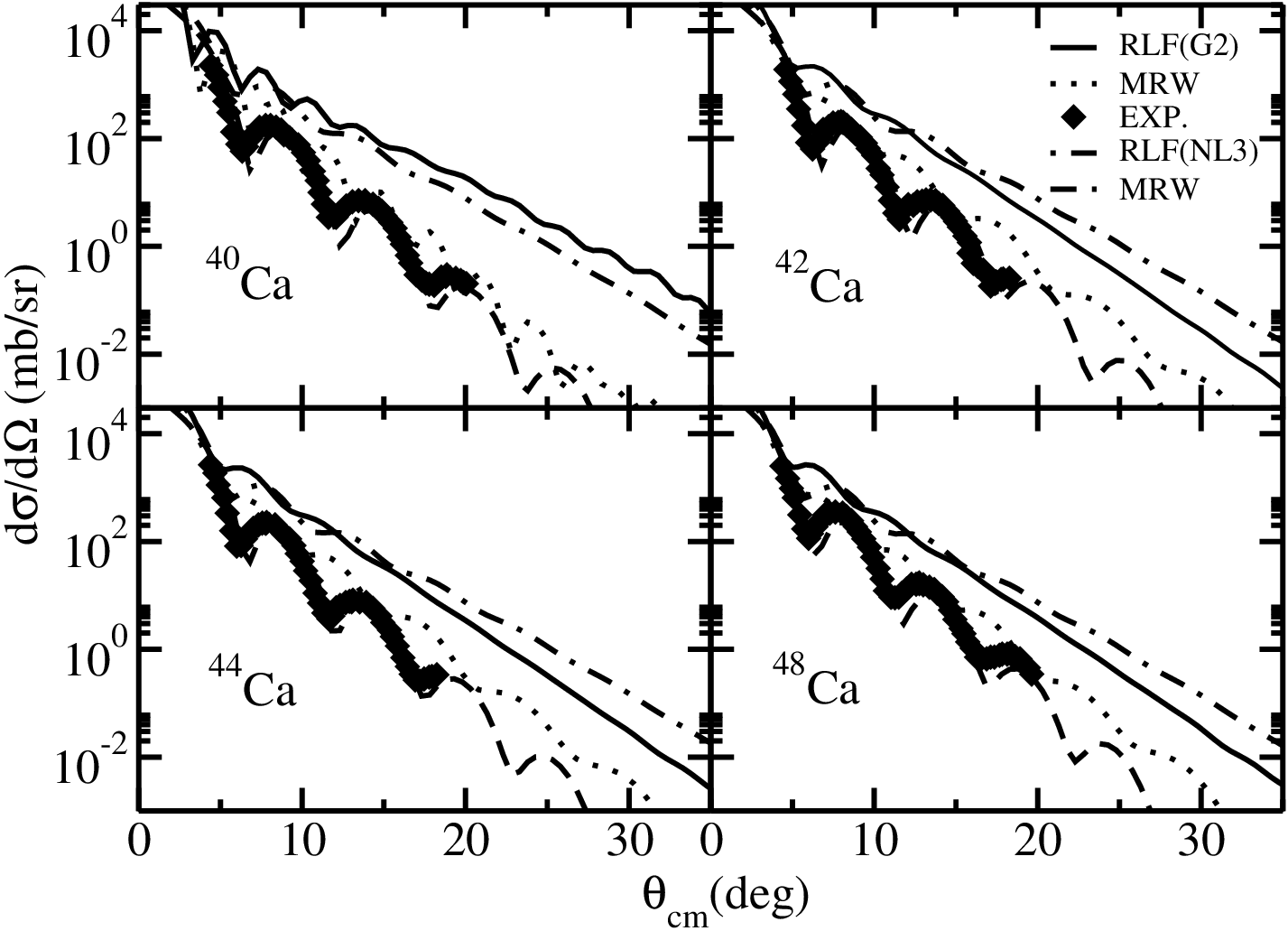}
\caption{{\it The elastic differential scattering cross-section
($\frac{d\sigma}{d\Omega}$) as a function of scattering angle
$\theta_{cm}$(deg) for $^{40,42,44,48}Ca$ using both
RLF and MRW parametrisations. The value of $\frac{d\sigma}{d\Omega}$
is shown for RMF (NL3) and E-RMF (G2) densities.
}}
\end{center}
\end{figure}

In Fig. 2., we have plotted the elastic scattering cross-section of 
the proton with $^{40,42,44,48}Ca$ at laboratory energy $E_{lab}=$1000 
MeV using both densities obtained in the NL3 and G2 parameter sets with 
RLF and MRW parametrisations. The experimental data \cite{expt78} are 
also given for comparison. It is reported in Refs. \cite{neil83,horo90} the
superiority of RLF over MRW for lower energy ($E_{lab}\leq 400$ MeV),
however the MRW shows better results at energy $E_{lab} > 400$ MeV.
In the present case, our incident energy is 1000 MeV which matches
better (MRW) with experimental values. This is consistent with the 
optical potential also (see Fig. 1). 
From the differential cross-section for both NL3 and G2 densities with 
MRW parametrization, it is clearly seen that $\frac{d\sigma}{d\Omega}$ 
with NL3 desity is more closer to experimental data which insist not 
only the importance of parametrization (RLF or MRW) but also to choose 
proper density input for the reaction dynamics.
Analysing the elastic differential cross-section along the isotopic chain
of Ca from A=40 to 48, the calculated results improve with increasing
mass number of the target.

\begin{figure}[ht]
\begin{center}
\includegraphics[width=1.0\columnwidth]{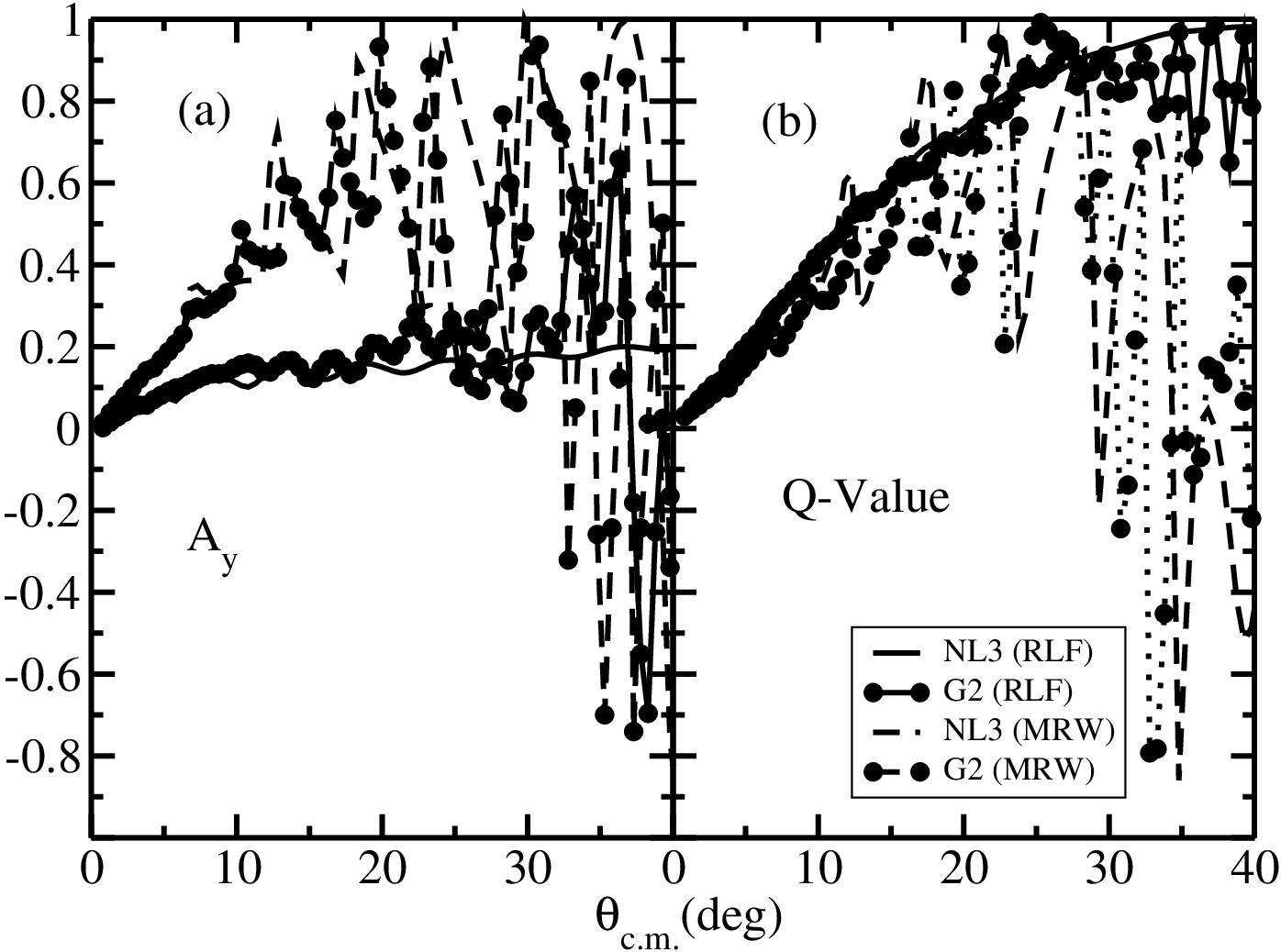}
\caption{{\it (a) The calculated values of analysing power $A_y$
as a function of scattering angle $\theta_{cm}$(deg) for $^{40}Ca$
(b) The spin observable $Q-$value as a function of scattering angle
$\theta_{cm}$(deg) for $^{40}Ca$. In both (a) and (b),
the RLF and MRW parametrisations are used with RMF (NL3) and
E-RMF (G2) densities.
}}
\end{center}
\end{figure}

The analysing power for $p+^{40}Ca$ composite system is calculated 
from the general formulae given in eqns. (4) and (5) and are shown in
Fig. 3 with RLF and MRW. The $A_y$ and $Q-$values obtained by NL3 and
G2 sets almost matches with each other both in RLF and MRW. But if we
compare the results with  RLF and MRW it differs significantly. Again,
we get a small oscillation of $A_y$ and $Q$ in G2 set with increasing 
scattering angle $\theta_{c.m.}^0$ for RLF which does not appear in 
NL3 set. There is a rotation of $Q-$value from positive to 
negative direction when we calculate with MRW parametrization,
which does not appear in case of RLF parametrization. This rotation 
shows a shining path towards the formation of exotic nuclei in the 
laboratory.  

In summary, we calculate the density distribution of protons and neutrons
for $^{40,42,44,48}Ca$ by using RMF (NL3)  and E-RMF (G2) parameter sets. 
We found similar density distribution for protons and neutrons in both
the sets with a small difference at the central region. This small difference
in densities make a significant influence in the prediction of 
optical potential, elastic differential cross-section, analysing power 
and the spin observable for $p+Ca$ systems.  The effect of kinematic 
parameters for reaction dynamics, RLF and MRW, are also highly sensitive 
to the predicted results. That means, the differential scattering 
cross-section and scattering observables are highly depent on the
input density and the choice of parametrisation.

\end{document}